\documentclass[11pt,a4paper]{article}
\usepackage{jheppub}

\usepackage{amsmath,amssymb}
\usepackage{verbatim,graphicx}

\usepackage{setspace}
\usepackage{color}
\usepackage[all]{xy}

\newcommand{\ket}[1]{| #1\rangle}
\newcommand{\bra}[1]{\langle #1|}

\DeclareMathOperator{\sign}{sgn}

\newcommand{\beq}{\begin{equation}}
\newcommand{\beqs}{\begin{equation*}}
\newcommand{\eeq}{\end{equation}}
\newcommand{\eeqs}{\end{equation*}}

\allowdisplaybreaks
\begin{document}
\setlength{\unitlength}{1mm}
\title{Open giant magnons on LLM geometries}

\author{David Berenstein,}
\author{Adolfo Holguin}
\affiliation { Department of Physics, University of California at Santa Barbara, CA 93106
}
\emailAdd{dberens@physics.ucsb.edu}
\emailAdd{adolfoholguin@physics.ucsb.edu}
\abstract{
We compute sigma model solutions for rigidly rotating open strings suspended between giant  gravitons in general LLM geometries.
These solutions are confined to the LLM plane.
These all have a dispersion relation for $\Delta-J$ that is consistent with saturation of a BPS bound of the centrally extended spin chain.
For the special case of circularly symmetric LLM geometries, we can further evaluate the amount of angular momentum $J$ carried
by these strings. This quantity diverges for string configurations that try to move between different ``coloring regions" in the LLM plane.
All of these quantities have a perturbative expansion in the t'Hooft coupling. For the strings suspended between AdS giants, we can compute in field theory the leading result of
$J$ carried by the string via an analytic continuation of the $SU(2)$ result, with the help of the Bethe Ansatz for the $SL(2)$ sector.
We thus provide additional information on how the radial direction of $AdS$ arises from (open) spin chain calculations.
}

\maketitle

\section{Introduction}

The AdS/CFT correspondence has established an equivalence between some theories of quantum gravity in asymptotically AdS spacetimes and certain gauge theories.
The most celebrated example is the equivalence between IIB string theory on $AdS_5\times S^5$ and ${\cal N}=4 $ SYM theory \cite{Maldacena:1997re}.

The free strings propagating on the $AdS_5\times S^5$ background are believed to be integrable for all values of the t'Hooft coupling. A review of the main results in this direction can be found in  \cite{Beisert:2010jr}. On the field theory side, the integrability takes the form of a spin chain Hamiltonian \cite{Minahan:2002ve,Beisert:2003yb}. The spin chain acts on the list of gauge invariant local operators, the states being generated by traces of words of local fields of ${\cal N}=4 $ SYM and their derivatives. Finding the energy spectrum on the spin chain side corresponds to computing the eigenvalues of the matrix of anomalous dimension of the operators in ${\cal N}=4 $ SYM.

 Integrability, combined with supersymmetry is very powerful. A particularly important result that combines the two is the dispersion relation for magnons on the gauge theory spin chain \cite{Beisert:2005tm}. It follows from
a central extension in the symmetry algebra of the spin chain and from the fact that magnons are in short representations of the centrally extended symmetry algebra.
This shortness condition fixes the kinematics.

When the magnons carry a lot of momentum on the spin chain, they become geometrically large string solutions in the AdS dual.  These are called giant magnons \cite{Hofman:2006xt}. These also carry central charge on the spin chain. The total central charge of a closed string state vanishes because of the level-matching constraint.
In the spin chain side this arises from the cyclic property  of the trace \cite{Berenstein:2002jq}.

The central charge on the gauge theory spin chain can also be sourced by open boundary conditions. These can be realized
 by supersymmetric D-branes in the AdS side, with open strings attached to them.
These D-brane states provide a very tractable connection between the gauge theory dynamics  and the $AdS$ geometry.
 This connection to the central charge extension on the spin chain and gravity dual side has been analyzed in the works \cite{Berenstein:2013md,Berenstein:2013eya,Berenstein:2014isa,Berenstein:2014zxa,Dzienkowski:2015zba,deCarvalho:2020pdp,Berenstein:2020grg}.
 Particularly, it has been suggested in \cite{Berenstein:2014zxa} that the central charge extension on the spin chain side is very closely related to the central charge extension of the Coulomb branch of ${\cal N}=4 $ SYM. Our recent work \cite{Berenstein:2020grg} showed how this works on the sigma model side for open string states suspended between D-brane states made of AdS giant gravitons. A complete picture in the analysis of the spin chain side is  still missing.

 As a reminder, giant gravitons are D-brane states that  preserve half the supersymmetry of the ${\cal N}=4 $ SYM theory. They can grow into the sphere \cite{McGreevy:2000cw}, or into the AdS  directions \cite{Grisaru:2000zn,Hashimoto:2000zp}.
 The ones that grow in the AdS directions are related to (classical)  spontaneous symmetry breaking from $U(N)\to U(N-1)\times U(1)$ via the Higgs mechanism, which generates expectation  values for the scalars \cite{Hashimoto:2000zp} (see also \cite{Berenstein:2004kk}).
 All of these D-brane states can be understood in terms of the classification of half-BPS states in ${\cal N}=4$ SYM in terms of Young tableaux \cite{Corley:2001zk}. Sphere giant gravitons are represented by long single columns \cite{Balasubramanian:2001nh}, while AdS giant gravitons are long single rows.

Very importantly, the AdS giant gravitons explore the radial direction of the $AdS_5$ geometry. This has always been the most mysterious emergent dimension in the AdS/CFT correspondence.
It has been related to a UV-IR relation \cite{Susskind:1998dq}, where the position in AdS is related to the UV scale of physics on the boundary.  The radial direction has also been related to the Renormalization group flow \cite{deBoer:1999tgo} and via the AdS D-branes, it is also related to the Higgs mechanism.

The radial direction on the spin chain side is much less well understood. Some states that explore the radial direction appear when rotating strings in AdS are studied \cite{Gubser:2002tv}, see also \cite{Frolov:2003qc,Kruczenski:2004wg}. They are characterized by logarithmic contributions in the spin quantum number to their anomalous dimension.  An argument for their logarithmic scaling of anomalous dimension is given in \cite{Alday:2007mf}.
The open strings strecthing between AdS giants that have been studied previously by us \cite{Berenstein:2020grg} do not have such logarithmic contributions to their anomalous dimension. Instead, their anomalous dimensions are governed by supersymmetry, and in particular, by the amount of central charge they carry. Their dispersion relation is
\begin{equation}
\Delta -J = \sqrt{Q^2 + \frac{\lambda}{4\pi^2} |{\cal Z}|^2},\label{eq:disp}
\end{equation}
where $Q$ is the angular momentum on $S^3\subset AdS_5$ and ${\cal Z}$ is the central charge of the open string, in geometric units.
At very large angular momentum on the sphere ($Q\to\infty$), for the giant magnons suspended between D-branes, their anomalous dimension can be arbitrarily close to zero, even at strong coupling.
This follows because the square root can be expanded in powers of ${\cal Z} /Q$. This is a power series in the t'Hooft coupling $\lambda$, and therefore one can in principle match coefficients order by order in perturbation theory on the CFT side. 
Other sets of works suggest that the giant magnon dispersion relation also plays a role in more general geometries. In particular, it has been argued that
the central charge extension controls magnon dispersion relations in concentric LLM geometries \cite{Chen:2007gh,Koch:2016jnm}. The LLM geometries are solutions of
type IIB supergravity on $AdS_5\times S^5$ that preserve exactly half of the supersymmetries \cite{Lin:2004nb}. They can be thought of as condensates of sphere giant gravitons
and/or AdS giant gravitons. In the field theory dual, these condensates are expected to describe the quantum corrected half-BPS Coulomb branch of the field theory, where the corrections are induced by the boundary sphere curvature. Because they describe part of the Coulomb branch, one would expect that the central charge extension of the Coulomb branch has a role to play in the dynamics.

The LLM  geometries depend on a two coloring of a degeneration plane (the LLM plane). If the coloring is made out of concentric circles, the background geometry has a well defined additional circular symmetry generated by a $U(1)$ R-charge $J$.  This is the same $J$ that appears in \eqref{eq:disp}. For a general LLM background, only $\Delta-J$ is well defined. The backgrounds
break the $J, \Delta$ symmetries independently, leaving $\Delta-J=0$ for the background configuration. The concentric geometries are eigenstates of $J,\Delta$ in the dual field theory.

The purpose of the current work is to address this idea on the sigma model side.  In particular, we want to understand exact solutions of the sigma model  for open strings stretching between sphere giant gravitons or AdS giant gravitons in general LLM geometries. These are more complicated  geometries than $AdS_5\times S^5$. To get a simple finite answer, there must be interesting cancellations taking place in the gravity calculation.
One of our goals to see how this analytic behavior arises in the string
sigma model computation.

In particular, we will find exact expansions in the t'Hooft coupling  for $\Delta-J$ as above, that can in principle be matched to perturbative computations in field theory.
It turns out, that even though the
sigma model in these geometries is not expected to be integrable (for example, a naive Bethe ansatz is expected to have inelastic boundary conditions \cite{Koch:2016qkp}), it is under enough analytic control so that these BPS strings are analytically solvable and the  dispersion relations for the open strings will look identical to equation \eqref{eq:disp}. We pay  extra attention to geometries that correspond to concentric circles, because they allow us to explore the amount
of  charge $J$ that the string carries.
It will turn out that the quantity $J$ contains additional information that is not carried by either the angular momentum $Q$ or the central charge. It depends on more details of the precise position of the string in the background geometry. Nevertheless, after some manipulations, we will show that it also gives rise to an expansion in the t'Hooft coupling that can be matched between  the spin chain and the geometry, at least to leading order.
These seem to be non-protected quantities associated to the symmetries that are spontaneously broken by the D-brane background
for these states. These determine in the end if a putative string state belongs to the Hilbert space or if it does not. If the $J$ charge diverges, the string state is not allowed. Such phenomena already arise in the spin chain computations \cite{Berenstein:2014zxa}, so it is important to understand their behavior in the gravity dual setup as well.

Another interesting aspect of the open strings between sphere giant gravitons is that there is a relation of the geometric sigma model solution and the Bethe ansatz on the spin chain
\cite{Berenstein:2014isa}. When one studies open strings attached to these giant gravitons, sites can ``jump in" and ``jump out" of the spin chain \cite{Berenstein:2005fa}. To have a more standard description, one realizes the spin chain in a bosonized language. One writes the states in terms of the number of sites between defects on the spin chain, rather than in terms of spin up and spin down state. Now, the number of sites in the spin chain is fixed, and the boundary conditions allow number non-conservation for the bosonic excitations instead.
If one writes coherent states for these generalized bosons, one finds that the equations that lead to the ground state of the spin chain can be understood as a bound state condition on the S-matrix of the magnons, subject to corresponding boundary conditions.
Similar results are not known in the dual $SL(2)$  sector. An approximation for the $SL(2)$ sector at very large vevs of the central charge for strings stretching between a dual giant graviton and itself can be found in \cite{Correa:2006yu}, where
the ``jumping in'' and ``jumping out'' of letters is self-consistently  ignored in the limit of large spin/central charge.
A similar connection to the Bethe ansatz is not known.

We provide evidence in this paper that the open strings stretching between dual giant gravitons also have an interpretation in terms of zeros of an S-matrix for the $SL(2)$ sector. In particular, we get a better understanding of the analytic continuation of the $SU(2)$ spin chain to the $SL(2)$ sector. We will also show that our interpretation of the analytic continuation is
compatible with the sigma model calculations.

The paper is organized as follows. In section 2 we review the form of all half BPS solutions to type IIB supergravity, and we re-express them in form that makes certain cancellations clearer. In section 3 we provide an explicit example of the open string solutions in question by considering the case of $AdS_5\times S^5$. In section 4 we solve for open strings stretching between both sphere and AdS giants in a general half BPS geometry, finding very similar expressions to those in the case of $AdS_5\times S^5$. In section 5 we concentrate on concentric half-BPS geometries, for which we study the form of the R-charge $J$ and its relation to the metric of the half-BPS geometry. In the case of $AdS_5\times S^5$, we study various limits for which this expression simplifies, and match the leading sigma model answer on $\mathbb{R}\times S^3$ to a computation on the dual one-loop $SU(2)$ spin chain. We are able to interpret the equations that lead to the sigma model solution as a continuum version of the condition for having a pole of the magnon S-matrix for the  $SU(2)$ and  $SL(2)$ sectors. The answers on both sectors are related to each other by an analytic continuation of the radial parameter of the LLM plane.

\section{Review of LLM Geometries}
The most general $\frac{1}{2}$-BPS solution to IIB 10d supergravity is given by the ansatz \cite{Lin:2004nb}:

\begin{equation}
\label{llm metric}
    ds^2= -\frac{y}{\sqrt{\frac{1}{4}-z^2}}(dt+ V)^2+ \frac{\sqrt{\frac{1}{4}-z^2}}{y}(dy^2+ dx_1^2+ dx_2^2)+ y \frac{\sqrt{\frac{1}{2}-z}}{\sqrt{\frac{1}{2}+z}}d\Omega_3^2+ y \frac{\sqrt{\frac{1}{2}+z}}{\sqrt{\frac{1}{2}-z}}d\Tilde{\Omega}_3^2
\end{equation}
The only free parameter of this metric is an auxiliary function $z$ of the coordinates $y,x_1, x_2$, which satisfies a six dimensional Laplace equation with rotational symmetry along four directions:
\begin{equation}
\begin{aligned}
    d \left(\star_3  \frac{dz}{y} \right)&=0\\
   y dV&= \star_3 dz
\end{aligned}
\end{equation}
Where $\star_3$ is the Hodge star operation on the coordinates $x_1, x_2,y$. In order to ensure the regularity of \eqref{llm metric}, we must have that the quantity $\mathcal{H}^{-2}= \frac{y}{\sqrt{\frac{1}{4}-z^2}}$ remains finite as $y$ approaches zero. This means that $z= \pm \frac{1}{2}$ on the $y=0$ plane which we will call the LLM plane. Because of this, the metric is completely determined by a coloring of the LLM plane into regions where $z= \pm \frac{1}{2}$. It will also  be convenient to rewrite the metric in a more compact form:

\begin{equation}
\label{llm_metric2}
    ds^2 = \mathcal{H}^{-2}\left( -(dt+V)^2+ \left(\frac{1}{2}-z\right)d\Omega_3^2 + \left(\frac{1}{2}+z\right)d\Tilde{\Omega}_3^2\right)+ \mathcal{H}^{2}\left(dy^2+ \delta_{ij}dx^idx^j \right)
\end{equation}
This is convenient since the parametrization \eqref{llm_metric2} makes the metric explicitly regular at $y=0$ inside the colored regions. Generically, at the boundary of a droplet the one form $V$ becomes singular, but such singularities can be eliminated via coordinate transformations.  An explicit form of $V$ is:

\begin{equation} \label{v solution}
    V_i(x_1,x_2,y)= \frac{\epsilon_{ij}}{2 \pi} \oint_{\partial \mathcal{D}}\frac{dx'_j}{(\bold{x}-\bold{x}')^2+y^2}
\end{equation}
Here the integration is taken along the boundaries of the droplets.
This guarantees that $V_i\to 0$ as $|x|,|y|\to \infty$.
Finally, the regions $z=\pm \frac{1}{2}$ are the degeneration loci of the either one of the two three-spheres, which is clear from equation \eqref{llm_metric2}.

\section{Open Strings on $AdS_3 \times S^1$}
Now we wish to review the solutions of the Nambu-Goto sigma model corresponding to rigidly rotating strings on a $AdS_3 \times S^1$ subspace of $AdS_5\times S^5$ that appeared in \cite{Berenstein:2020grg}, but studying the solution directly in the LLM cordinates instead. The corresponding metric in the coordinates \eqref{llm metric} is described by a single droplet configuration on the LLM plane of radius $r_0$:

\begin{equation}
\begin{aligned}
    z(r,y;r_0)&= \frac{r^2-r_0^2+ y^2}{2\sqrt{(r^2+r_0^2+y^2)^2-4r^2r_0^2}}\\
    V_{\phi}&= -\frac{1}{2}\left( \frac{r^2+r_0^2+ y^2}{\sqrt{(r^2+r_0^2+y^2)^2-4r^2r_0^2}}-1\right)
    \end{aligned}
\end{equation}
We will be interested in solutions that reside at the $y=0$ locus, with a D3 brane wrapping the non-vanishing three-sphere and with the string rotating along a circle of the non-vanishing sphere. The effective metric for the space on which the strings move can be written in the form:
\begin{equation}
   ds^2= - s \left(1-\frac{r^2}{r_0^2} \right)(dt + V_\phi d\phi)^2+ s\frac{\left( dr^2+ r^2 d\phi^2 \right)}{\left( 1-\frac{r^2}{r_0^2}\right)} + s \left(1-\frac{r^2}{r_0^2} \right)\left(\frac{1}{2}(1- s)d\psi^2+ \frac{1}{2}(1+ s)d\tilde{\theta}^2 \right)
\end{equation}
with $s= \sign(r_0-r)$.
The effective geometry for $r>r_0$ corresponds to $AdS_5 \times S^1$, while $r<r_0$ corresponds to  $\mathbb{R}_t\times S^5$. One should also note that the behavior of $V_\phi$ at $y=0$ is non-trivial as  $r$ crosses $r_0$:
\begin{equation}\label{V}
    \begin{aligned}
    V_\phi(r<r_0, y=0) &= \frac{r^2}{r_0^2-r^2}\\
     V_\phi(r>r_0, y=0) &= \frac{r_0^2}{r^2-r_0^2}
    \end{aligned}
\end{equation}


For $r>r_0$, the metric is
\begin{equation}
\label{ads3 metric}
   \frac{ds^2}{r_0}= -\left(r^2-1 \right)dt^2+ 2  dt d\phi +\frac{ dr^2 }{\left(r^2-1  \right)}+   d\phi^2 + (r^2-1) d\psi^2
\end{equation}
Where the variable $r$ has been re-scaled to be unitless. Now we can consider the string sigma model on this geometry, concentrating on rigid open string solutions that end on two static dual giant gravitons. The boundary conditions allow for the endpoints of the string to move freely along the $\psi$ direction, so we restrict to configurations where the string endpoints co-rotate at the same angular velocity $\beta$. A convenient ansatz for the embedding coordinates is of the form:
\begin{equation} \label{ansatz1}
    \begin{aligned}
    t= \tau\\
    r= r(\sigma)\\
    \phi= \phi(\sigma)\\
    \psi= \beta \tau + g(\sigma)
    \end{aligned}
\end{equation}
Then, the string action in these coordinates is given by:
\begin{equation}
    S= -\frac{\sqrt{\lambda}}{2\pi} \int d\tau d\sigma \sqrt{(r^2-1)^2 g'^2 + 2 \beta (r^2-1)\phi' g' + \phi'^2(\beta^2(1-r^2) + r^2) +(1-\beta^2)r'^2 )}
\end{equation}
For the coordinate $g$ one has to impose Neumann boundary conditions, which is equivalent to saying that the worldsheet current density $\frac{\partial \mathcal{L}}{\partial g' }$ vanishes at the end points of the string. In addition to this, since the action is independent of $g$ this current must vanish identically along the string. This leads to the condition:

\begin{equation}
\label{ads3 neumann}
    g'=-\frac{\beta \phi'}{r^2-1}
\end{equation}
which can be used to eliminate $g$ in the action. This reduces the problem to a geodesic equation on a flat plane, as long as $r\neq 1$ where equation \eqref{ads3 neumann} degenerates:

\begin{equation} \label{ads string}
    S =-\frac{\sqrt{\lambda}}{2\pi}  \sqrt{(1-\beta^2)} \int d\tau d \sigma \sqrt{r^2 \phi'^2 + r'^2}
\end{equation}
Due to the rotational symmetry of the droplet, a general solution can always be transformed to one determined by a pair of angles $\phi_1, \phi_2$ from the $x_1$ axis and the closest approach to the origin $a$. These are the same solutions studied in \cite{Berenstein:2020grg} in slightly different coordinates.  In particular, the conserved charge associated to time translations of the coordinates \eqref{ads3 metric} follows a relativistic dispersion relation

\begin{equation}\label{charges}
\begin{aligned}
\epsilon &= \sqrt{Q^2+ \frac{\lambda}{4 \pi^2}|\mathcal{Z}|^2}\\
Q&= \frac{\sqrt{\lambda}}{2 \pi} \frac{\beta |\mathcal{Z}|}{\sqrt{1-\beta^2}}\\
|\mathcal{Z}|&=  \int_{\phi_i}^{\phi_f} d\phi \sqrt{r^2+ \left(\frac{d r}{d\phi} \right)^2}
\end{aligned}
\end{equation}
where $Q$ is the angular momentum associated to rotations along $\psi$ and $\mathcal{Z}$ is the central charge associated to the separation of the branes. It is also important to notice that the density of central charge and angular momentum per unit length are constant along the string. One can also check that the angular momentum density $J$ associated to rotations along the LLM plane diverges if the string solution touches boundary of the droplet. We will make this more explicit in section 5.
\section{Open Strings on LLM Geometries}
We can now discuss more general solutions corresponding to open strings on general $\frac{1}{2}$-BPS geometries. As we will see these share many similarities to the solutions discussed in the previous section.
\subsection{Strings outside a droplet}
First we consider the case for a string inside a connected region with $z= -\frac{1}{2}$. Within each of these regions we have a non-vanishing three-sphere on which we can wrap D3 branes. In the end, we will be interested in rotating string solutions, so we will also single out a circle within this three-sphere with coordinate $\psi$ on which the string endpoints rotate. The branes will sit at $y=0$, but it is convenient to keep the value of $y$ unfixed along the string as this makes the various cancellations clear. The appropriate ansatz for the embedding coordinates is the similar to the one before,

\begin{equation}
    \begin{aligned}
    t&= \omega \tau\\
    x_i&= x_i(\sigma)\\
    y&= y(\sigma)\\
    \psi&= \beta \tau + g(\sigma)\\
    \end{aligned}
\end{equation}
except that the effective metric is now of the general form:
\begin{equation}
\label{llm metric3}
    ds^2 = \mathcal{H}^{-2} \left( -(dt+V)^2+ \left(\frac{1}{2}-z\right)d\psi^2 \right)+ \mathcal{H}^{2}\left(dy^2+ \delta_{ij}dx^idx^j \right)
\end{equation}
The Nambu-Goto action in these coordinates is:
\begin{equation}\label{ng llm}
\begin{aligned}
    S &= \frac{\sqrt{\lambda}}{2 \pi}\int d\tau d \sigma \sqrt{\mathcal{G}}\\
    \mathcal{G} &= \mathcal{H}^{-4} \bigg( (\frac{1}{2}-z)g'^2 + 2(z-\frac{1}{2})g' V_ix_i'\\ &+ \beta^2(z-\frac{1}{2})(V_ix_i')^2 \bigg)+ (1- \beta^2 (\frac{1}{2}-z))(x_1'^2+x_2'^2+ y'^2)
    \end{aligned}
\end{equation}
Where we have set $\omega=1$ for simplicity. The coordinate $g'$ can be eliminated by a combination of its equation of motion and boundary conditions as before. This leads to the simple relation which generalizes \eqref{ads3 neumann}:

\begin{equation} \label{llm neumann}
    g'= \beta V_ix_i'
\end{equation}
Once again, the relation \eqref{llm neumann} shows that the variable $g'$ becomes ill-defined whenever the string touches the boundary of a droplet \eqref{v solution} at $y=0$.  One can also express this relation in a way that is independent of the parametrization,
\begin{equation}
    dg= \beta V
\end{equation}
so that $dg$ is well defined in regions where $z$ is locally constant. Substituting this relation into the action \eqref{ng llm} will cancel the terms in $\mathcal{G}$ which are multiplied by the warp factor $\mathcal{H}$, which simplifies the action to the form:

\begin{equation} \label{action}
\begin{aligned}
     S &= -\frac{\sqrt{\lambda}}{2 \pi}\int d\tau d \sigma \sqrt{(1- \beta^2 (\frac{1}{2}-z))(x_1'^2+x_2'^2+ y'^2)}
       \end{aligned}
\end{equation}
One can also find a similar expression for the energy of the string by varying with respect to $\omega$:
\begin{equation}
\begin{aligned}
    \epsilon &= \frac{\sqrt{\lambda}}{2 \pi}\int d\tau d \sigma \sqrt{\frac{1}{(1- \beta^2 (\frac{1}{2}-z))}(x_1'^2+x_2'^2+ y'^2)}
    \end{aligned}
\end{equation}
In general, having the string extend in the $y$ direction makes the equations non-linear, but such configurations happen to not have minimal energies. The minimal energy configurations are those for which $y=0$ along the string, for which the action reduces to a geodesic problem on the LLM plane.
\begin{equation}
    S=- \frac{\sqrt{\lambda}}{2\pi} \int d\tau \sqrt{(1-\beta^2)(dx_1^2+ dx_2^2)}
\end{equation}
This is the same result as for open strings between D-branes in $AdS_5\times S^5$, so the kinematic features are the same.
As a result, this class of solutions also admit a giant magnon dispersion relation:

\begin{equation}
\begin{aligned}
    \epsilon &= \sqrt{Q^2+ \frac{\lambda}{4 \pi^2} |\mathcal{Z}|^2}\\
    |\mathcal{Z}|&=\int \sqrt{dx_1^2+ dx_2^2}
\end{aligned}
\end{equation}
Generically, an LLM geometry will not have rotational invariance along the the LLM plane due to the placement of sources for $z$. This means the charge $J$ associated to this rotation is no longer a good quantum number in the dual description. However, there is always an approximate translational symmetry in the limit that one zooms into the boundary of a droplet. The effective geometry in this limit is always a plane-wave, and the density of the momentum associated to the approximate translational symmetry will generically diverge. This is because such quantities are always proportional to the gauge potential $V$ which is not well defined at the interfaces between the different values of $z$.
\subsection{Inside a droplet}
The analysis for connected regions with $z=\frac{1}{2}$ is completely analogous to the one in the previous section. In this case there is a different non-vanishing three-sphere $\tilde{S^3}$, from which we single out a circle $\tilde{\theta}$.
The effective metric is a simple variation of \eqref{llm metric3}

\begin{equation}
    ds^2 = \mathcal{H}^{-2} \left( -(dt+V)^2+ \left(\frac{1}{2}+z\right)d\tilde{\theta}^2 \right)+ \mathcal{H}^{2}\left(dy^2+ \delta_{ij}dx^idx^j \right)
\end{equation}
The appropriate ansatz in this case is the same as before \eqref{ansatz1}, but we replace the variable $\psi$ by:
\begin{equation}
    \tilde{\theta}= \tilde{\beta} \tau + h(\sigma)
\end{equation}
The computation of the action is entirely analogous as to the discussion in the previous section. The analogous condition \eqref{llm neumann} that arises from the boundary conditions for $h$ is:
\begin{equation}
    h'= \tilde{\beta} V_i x_i'
\end{equation}
which tells us that $h'$ is the pullback of $V$ on the worldsheet inside the droplet regions. This means that $h'$ has the same singularities along interfaces as $g'$ did, so that continuing the variable $g$ to regions inside a droplet becomes problematic. After eliminating $h$, the action takes the same form as before with the appropriate change in kinematic factors:
\begin{equation}
    S=- \frac{\sqrt{\lambda}}{2\pi} \int d\tau \sqrt{(1-\tilde{\beta}^2)(dx_1^2+ dx_2^2)}
\end{equation}
Similarly, the energy can be easily shown to satisfy a similar relativistic dispersion relation
\begin{equation}
\begin{aligned}
\epsilon &= \sqrt{\tilde{Q}^2+ \frac{\lambda}{4 \pi^2}|\mathcal{Z}|^2}
\end{aligned}
\end{equation}
Where $\tilde{Q}$ is the angular momentum along the circle $\tilde{\theta}$.
\section{On-Shell Charges}
For this section we will concentrate on $\frac{1}{2}$-BPS geometries that correspond to concentric droplets and rings on the LLM plane. This is useful since we will want to study the behavior of the charge $J$ associated to rotations around the origin of the LLM plane. One important point that should be noted is that the coordinates \eqref{llm metric} are implicitly rotating with respect to an observer that is far away from the sources to whom the geometry looks like $AdS_5\times S^5$. So solutions that are static in these coordinates correspond to strings that rotate along a cycle that asymptotically looks like the equator of $S^5$. As such the charge $\epsilon$ associated to time translation symmetry in the LLM coordinates actually corresponds to $\Delta- J$ in the global $AdS$ coordinates. One can find the expression for $J$ for a general concentric concentric geometry by modifying the ansatz for the coordinate $\phi$ \eqref{ansatz1} to include time dependence,
\begin{equation}
    \phi= \varphi(\sigma) + \gamma \tau
\end{equation}
and in the end substituting the on-shell value $\gamma=0$. Unlike the charges $\epsilon, \mathcal{Z}, Q (\tilde{Q})$ the angular momentum $J$ turns out to be sensitive to the details of the geometry. This is because the general form of $\epsilon$ is fixed by supersymmetry \cite{Beisert:2005tm}, and the other charges assemble into a relativistic dispersion relation.
As discussed in \cite{Berenstein:2014zxa}, the shortening condition is essential to get the right multiplicities for light  open strings between nearby giants. This is what guarantees that the local physics looks like ${\cal N}=4$ SYM on the Coulomb branch.
 For concreteness we first concentrate on the case where the strings live in a region outside a single circular droplet on the LLM plane, and then we show that the analysis extends to solutions sitting inside the droplet. The resulting expression for $J$ and its density along the string are:

\begin{equation}\label{J}
\begin{aligned}
J&= \frac{\sqrt{\lambda}}{2 \pi} \int d\sigma \mathcal{J}\\
    \mathcal{J}&=\frac{V_\phi \left(\left(\phi '\right)^2 \left(r^2
   \left(\frac{\left(\mathcal{H}^2+1\right) \left(1-\beta ^2
  \right)}{2\mathcal{H}^2}+1\right)+\frac{\left(\frac{1}{\mathcal{H}^4}+\frac{1}{\mathcal{H}^2}+2\right) V_\phi^2 \left(\beta ^2 -1\right)}{2\mathcal{H}^2}\right)+ \left(r'\right)^2\right)}{ \sqrt{\left(1-\beta ^2
  \right) \left(r^2 \left(\phi
   '\right)^2+\left(r'\right)^2\right)}}
   \end{aligned}
\end{equation}

From this we can see that the density $\mathcal{J}$ is proportional to $V_\phi$, so that it becomes infinite at the boundaries of the droplets as claimed. Since the central charge density is constant on-shell we can also express this as:

\begin{equation}
    \begin{aligned}
       J&=
   \epsilon \int d\sigma V_\phi+ \left(\frac{\lambda}{4 \pi^2}\right)\frac{1}{\epsilon} \int d\sigma  \phi'^2\left( r^2(1+\mathcal{H}^{-2})-V_\phi^2 \left(\mathcal{H}^{-2}+ \frac{\mathcal{H}^{-6}+\mathcal{H}^{-4}}{2}\right)\right)\\
   \epsilon&=\sqrt{Q^2+ \frac{\lambda}{4 \pi^2}\mathcal{Z}^2}
    \end{aligned}
\end{equation}
One can also express the density $\mathcal{J}$  in terms of the quantum numbers $Q$, $\mathcal{Z}$, the t'Hooft coupling $\lambda$ and $r$. For completeness we will carry this out for the simplest droplet configuration studied in section 3, in the regime that the strings are far away from the droplet.
\subsection{Strings near the boundary}
We would like to evaluate \eqref{J} for a string solution that sits very far away from a single droplet, while the size of the string remains finite but large. These correspond to strings that sit near the boundary of $AdS_5$.  As it turns out, the expression \eqref{J} is not the same as the angular momentum measured by an asymptotic observer in global $AdS$, since the coordinates \eqref{ads3 metric} describe a rotating frame $\tilde{\phi}= \phi - t$, so that the expression for $J$ is actually a linear combination of the scaling dimension $\Delta$ and the spin $\tilde{J}$ seen by a static observer near the boundary. The two quantities $J,\tilde{J}$ are related by a simple change of coordinates, but it is more convenient to work with the expressions in \cite{Berenstein:2020grg} which have a clearer physical interpretation. More concretely, this is the choice of coordinates for which the scaling dimension $\Delta$ grows with the distance from the origin, while the spin $\tilde{J}$ becomes smaller:

\begin{equation} \label{adsJ}
    \tilde{J}=  \frac{\sqrt{\lambda}}{2\pi}\int d \tilde{\phi} \frac{(\frac{dr}{d \tilde{\phi}})^2+ \beta^2-1+ r^2}{(r^2-1)\sqrt{(1-\beta^2)(r^2+(\frac{dr}{d \tilde{\phi}})^2 )}}
\end{equation}
 Here $\tilde{\phi}= \phi - t$ corresponds to the coordinate for the equator of the $S^5$. We should also note that this expression for $\tilde{J}$ is not gauge invariant, as it will turn out that this expression is related to the one form $V$ that appears in the metric. On the one hand these gauge transformations can be always absorbed into a redefinition of the string coordinates $g$ and $h$ \eqref{llm neumann}. However,  one should also keep in mind that gauge transformations that vanish at infinity do not change the asymptotics of $\tilde{J}$, so that the coordinate choices for which $\tilde{J}$ vanishes at infinity are well-defined. In order to fix the residual gauge symmetry one has to choose coordinates that look asymptotically like static global $AdS$ rather than the rotating LLM coordinates. It will also be convenient to make the choice $V_r=0$ in order to keep the rotational symmetry explicit. \\ We wish to find an expression for \eqref{adsJ} in terms of the angular momentum $Q$, the string end points $\xi_1, \xi_2$, and the t'Hooft coupling $\lambda$. For this, it is best to re-express the integral using an affine parametrization for the the complex coordinate on the LLM plane $z= \xi_1 (1-s)+ \xi_2 s$. In order to eliminate the angular velocity $\beta$, we have to impose a double scaling limit based on \eqref{charges}. The particular double scaling limit we will be interested in comes from fixing the angular momentum $Q$ and the positions of the end points of the string:
\begin{equation}
    \frac{Q}{\mathcal{Z}}= \frac{\sqrt{\lambda}}{2 \pi }\frac{\beta}{\sqrt{1-\beta^2}}<\infty
\end{equation}
This is a physical choice of scaling, since the strings become tensionless in the relativitic limit $\beta \rightarrow 1$.
This leaves us with two independent parameters to tune which we can choose to be the ratio $\frac{Q}{\mathcal{Z}}$ and the t'Hooft coupling, since changing the value of $\beta$ has to be compensated by a change of $\lambda$ in order to keep $Q$ and $\mathcal{Z}$ fixed. This means that the angular velocity and the t'Hooft coupling can't be changed independently from each other.
The full expression for $\tilde{J}$ in terms of these parameters can be expressed as a sum of two terms:

\begin{equation}\label{tildeJ}
    \tilde{J}= \sqrt{Q^2+ \frac{\lambda}{4 \pi^2}|\xi_1-\xi_2|^2} \int_0^1 \frac{ds}{z \Bar{z}-1}+ \left(\frac{\lambda}{4 \pi^2}\right)\frac{1}{\sqrt{Q^2+ \frac{\lambda}{4 \pi^2}|\xi_1-\xi_2|^2}} \mathfrak{Im}\int_0^1 ds \frac{z \frac{d \Bar{z}}{ds}}{z\Bar{z}(z\Bar{z}-1)}
\end{equation}
This expression is interesting as it is a series expansion in $\lambda$ around zero, while also showing a scale separation given at small and large energies $\epsilon$.

In particular, one can expect that this quantity can be recovered via a perturbative calculation in the dual field theory since we have a series expansion in the t'Hooft coupling. This is different from in the other giant magnon solutions studied in the literature which always have infinite spin $\tilde{J}$ and correspond to closed strings \cite{Hofman:2006xt, Minahan:2006bd}. Alternatively, one can expand in $\kappa= \lambda/Q^2$, which can be done even at strong coupling. This is similar to how in the plane wave limit the
effective perturbation parameter depends on the quantum numbers of an excitation \cite{Berenstein:2002jq}.

We will reproduce the leading term of the expansion in the $\mathfrak{su}(2)$ sector by an explicit computation from the one-loop spin chain Hamiltonian with boundary conditions. Motivated by an analytic continuation of the Bethe ansatz, we will obtain an expression for the $\mathfrak{sl}(2)$ spin chain with open boundary conditions, even when we do not know the form of the precise computation on the dual field theory for this sector. In general, these boundary conditions are expected to break integrability, but the existence of a Bethe-like ansatz for the $\mathfrak{su}(2)$ in terms of Cuntz oscillator coherent states suggests that a similar story exists for the $\mathfrak{sl}(2)$ spin chain.  One should also note that even though the leading term is independent of $\lambda$, the computation requires knowing the one-loop mixing Hamiltonian for the $\mathfrak{su}(2)$ sector, and higher order corrections arise from higher loop contributions to the mixing of operators.
\subsubsection{Rapidly Rotating Strings}
An interesting limit to consider is when the strings rotate with a large angular velocity $\beta \rightarrow 1$ with fixed angular momentum $Q$. This is the limit where the t'Hooft coupling becomes small, so that the second term in \eqref{tildeJ} can be ignored.

\begin{equation}\label{integral}
    \tilde{J}= |Q| \int_0^1 \frac{ds}{| \xi_1 (1-s)+ \xi_2 s|^2-1}+ \dots
\end{equation}

The integral in \eqref{integral} is somewhat reminiscent of a Feynman parametrization, and can be evaluated explicitly:

\begin{equation}\label{Jintegral}
\begin{aligned}
    \tilde{J}&=  |Q| \frac{\arctan\left(\frac{|\xi_2|^2-|\xi_1|^2+|\xi_1-\xi_2|^2}{2\sqrt{|\xi_1 \times \xi_2|^2-|\xi_1-\xi_2|^2}}\right)-\arctan\left(\frac{|\xi_2|^2-|\xi_1|^2-|\xi_1-\xi_2|^2}{2\sqrt{|\xi_1 \times \xi_2|^2-|\xi_1-\xi_2|^2}}\right)}{\sqrt{|\xi_1 \times \xi_2|^2-|\xi_1-\xi_2|^2}}+ \dots\\
    \xi_1\times \xi_2&= \mathfrak{Im}\left(\xi_1\Bar{\xi_2} \right)= |\xi_1||\xi_2|\sin{\theta_{12}}
    \end{aligned}
\end{equation}
In the limit that the string end points are very far away from the origin we can ignore the 1 in the denominator of \eqref{integral}:

\begin{equation}\label{Jtilde2}
    \tilde{J}=  \frac{ \theta_{12}|Q|}{|\xi_1| |\xi_2| \sin\theta_{12}} +\dots
\end{equation}
Although the expression \eqref{Jtilde2} is regular at $\theta_{12}=0$ where the string end points are colinear, there is a divergence at $\theta_{12}=\pi$ coming from the fact that the string has to cross the droplet.
In the strict $\xi\rightarrow\infty$ limit, the leading order contribution for $\tilde{J}$ vanishes. Generically, in the strict $\beta \rightarrow 1$ limit the divergent contributions to the LLM angular momentum will decouple so that the expressions for $\tilde{J}$ and $J$ match. More explicitly, the expression \eqref{J} becomes much simpler in this limit.

\begin{equation} \label{leadingJ}
    J = \frac{\sqrt{\lambda} \mathcal{Z}}{\sqrt{1-\beta^2}} \int_0^1 d s V_\phi\left(r(s), y=0\right)+ \dots
\end{equation}
Where $r(s)$ is an affine parametrization for the string in the LLM plane. It also turns out that the expression \eqref{leadingJ} is valid inside and outside the droplet regions as long as one chooses the correct branch for $V_\phi(y=0)$.
Another thing to note is that value of $V_\phi$ inside of the droplet is related to its value outside the droplet by a change of sign and a transformation $r\rightarrow \frac{r_0^2}{r}$, where $r_0$ is related to the $AdS$ radius $r_{AdS}^2= r_0$. By restoring the dependance on $r_0$, the expression \eqref{Jtilde2} should be understood as the leading order expansion in $r_0/m^2$, where $m^2$ is a large mass parameter compared to $r_0$. This can be done by either sending the string end points to infinity, or by considering a small droplet. This is also the regime where the contribution to masses of the conformally coupled scalars coming from the curvature of $\mathbb{R}\times S^3$ is negligible in the field theory, which is a decompactification limit of the $S^3$. This suggests that the leading non-vanishing term in  $J$ at large $\beta$ should be reproducible from a Coulomb branch computation, while the higher order terms in powers $r_0$ should come from taking into account correctly the mixing between the higgsinos and gauginos, since a priori the massive vectors do not couple to the curvature $R_{S^3}\sim R_{AdS}$. A calculation with background fields properly included would look similar to \cite{Berenstein:2007zf}, where the localization in the geometry is provided directly by the D-brane background fields, rather than a saddle point.

For more general concentric droplet geometries the expression for $V_\phi$ outside the largest droplet is given by a linear combination of droplets \cite{Lin:2004nb}:

\begin{equation}
    V_\phi(r, y=0)= \sum_{i=0}^k \frac{ (-1)^{i}r_i^2}{r^2-r_i^2}
\end{equation}
The leading expression for $\tilde{J}$ in this case can be easily seen to come from adding the contributions coming from all the droplets and holes:

\begin{equation}
\begin{aligned}
       \tilde{J}&= |Q| \sum_{i=0}^k (-1)^i \tilde{J_i}\\
       \tilde{J_i}&= r_i^2 \left( \frac{\arctan\left(\frac{|\xi_2|^2-|\xi_1|^2+|\xi_1-\xi_2|^2}{2\sqrt{|\xi_1 \times \xi_2|^2-r_i^2|\xi_1-\xi_2|^2}}\right)-\arctan\left(\frac{|\xi_2|^2-|\xi_1|^2-|\xi_1-\xi_2|^2}{2\sqrt{|\xi_1 \times \xi_2|^2-r_i^2|\xi_1-\xi_2|^2}}\right)}{\sqrt{|\xi_1 \times \xi_2|^2-r_i^2|\xi_1-\xi_2|^2}}\right)+ \dots
\end{aligned}
\end{equation}
This suggests that the leading order computation on the dual field theory side also comes from summing simple contributions and at leading order the droplets don't affect each other.
 \subsection{Strings Inside a Droplet}
 We can also do the analogous computation for string solutions sitting inside a circular droplet region. In this case the motion of the string is restricted to an $S^3\times \mathbb{R}$ subspace of $AdS_5\times S^5$. It is well known that the giant magnons on $S^3\times \mathbb{R}$ have a dual description in terms of a $\mathfrak{su}(2)$ integrable spin chain whose Hamiltonian computes the mixing of operators.

 \subsubsection{Dual Spin Chain Picture}
 The one loop Hamiltonian for the $SU(2)$ sector with open boundary conditions is of the form \cite{Berenstein:2005fa,Berenstein:2014zxa}:

 \begin{equation}
     H_1= \lambda \sum_{i=0}^k (a^\dagger_i-a^\dagger_{i+1} )(a_i-a_{i+1})
 \end{equation}
 Here $a_0$ and $a_{k+1}$ are complex numbers describing the collective coordinates of the giant gravitons, and the $a_i$ are Cuntz oscillators satisfying the algebra:
 \begin{equation}
 \begin{aligned}
     a_ia^\dagger_j&= \delta_{ij}\\
     a_i^\dagger a_i&= 1- \ket{0}_i\bra{0}_i
     \end{aligned}
 \end{equation}
 The integer $k+1$ is associated to the angular momentum $\tilde{Q}$ which counts the number of sites of the oscillator chain, which has a length corresponding to the central charge $\mathcal{Z}$. In the gauge theory variables $k$ counts the number of $Y$ insertions between $Z$ in the operator:
 \begin{equation}
\begin{aligned}
      \mathcal{O}\sim& \dots ZZZY^{L_1}Z\dots Z Y^{L_n}\dots  \\
      \sum_i L_i& =k+1
\end{aligned}
 \end{equation}

 A complete combinatoric picture of how the strings are attached to the giants and how the boundary conditions emerge is found in the works
 \cite{Balasubramanian:2004nb,deMelloKoch:2007rqf,deMelloKoch:2007nbd,Bekker:2007ea}. To fix the angular position of the brane one needs to
add a coherent state description of the D-branes \cite{Berenstein:2013md}.
 These discussions usually only pertain to the $SU(2)$ sector.
For he $SL(2)$ sector,  an incomplete description in the Cuntz oscillator language is found in \cite{Correa:2006yu}, which was derived from \cite{Beisert:2003jj}.
This description of the $SL(2)$ sector is not that of a  spin chain with local nearest neighbor terms only. This makes a direct analysis very cumbersome.
When we discuss such calculations, we will sidestep this direct route of computation by utilizing ideas from the Bethe ansatz.

 We then proceed by considering an unnormalized coherent state for each oscillator:
 \begin{equation}
     \ket{z_i}= \sum_{n=0}^\infty z_i^n \ket{n}
 \end{equation}
 Substituting this into the Hamiltonian and minizing the energy one obtains the condition $z_i-z_{i+1}=\delta\mathcal{Z} $ for every adjacent pair of sites, where $\delta\mathcal{Z}$ is a constant which acts as a lattice spacing for the string. The quantity $\delta {\cal Z}$ is generically complex, but we can always align the coordinates so that it is real.

 This means that the central charge density along the chain $\frac{\mathcal{Z}}{\tilde{Q}}$ is a fixed constant. It can also be checked that this is in fact an eigenstate of the one-loop Hamiltonian with minimal energy. \\
 One can easily check that,
 \begin{equation}
\begin{aligned}
   \bra{\Bar{z}} z\partial_z \ket{z}&= \sum_{n=0}^\infty n z^n \ket{n}= \frac{z\Bar{z}}{(1-z\Bar{z})^2}\\
    \langle\Bar{z}|z\rangle&= \frac{1}{1-z\Bar{z}}
    \end{aligned}
    \end{equation}
    So that the average occupation number for each site is given by:
    \begin{equation}
        \frac{\bra{\Bar{z}} z\partial_z \ket{z}}{\langle\Bar{z}|z\rangle}= \frac{z\Bar{z}}{1- z\Bar{z}}
    \end{equation}
This occupation number also computes the R-charge $J$ for each oscillator, so that in total we have:

\begin{equation}\label{Rcharge}
 J=  \sum_{i=1}^k  \frac{z_i\Bar{z}_i}{1- z_i\Bar{z}_i}
\end{equation}
Since the central charge density along the string is constant, we can multiply each term by the central charge density $|\delta\mathcal{Z}|= |z_i-z_{i-1}|= \frac{\mathcal{Z}}{k}$
\begin{equation}
    J = \frac{|\tilde{Q}-1|}{\mathcal{Z}} \sum_{i=1}^k \frac{z_i\Bar{z}_i}{1- z_i\Bar{z}_i} \delta\mathcal{Z} \longrightarrow_{\delta\mathcal{Z}\rightarrow 0} |\tilde{Q}| \int_0^1 ds \frac{z\Bar{z}}{1- z\Bar{z}}
\end{equation}
The sum can then be approximated by an integral as we take the effective lattice spacing $\delta\mathcal{Z}$ to zero, by which one expects to recover the continuum string description.

\subsubsection{Sigma Model Computation}
The computation of the spin $J$ for the sigma model is straightforward. In this case we should take the metric with $r<1$ where the string motion occurs on $S^3 \times \mathbb{R}$ as opposed to $AdS_3 \times S^1$.  As before we will be interested in the double scaling limit that arises from taking $\tilde{\beta}\rightarrow1$ while holding the angular momentum $\tilde{Q}$ and central charge constant. The charge $\tilde{Q}$ and angular frequency $\tilde{\beta}$ should not be confused with $Q$ and $\beta$ although their roles are very similar. The leading order expression is simply:

\begin{equation}
    J= |\tilde{Q}|\int_0^1 ds V_\phi\left(r(s),y=0\right)
\end{equation}
Again, we can introduce an affine parametrization for the complex variable on the LLM plane $z= \eta_1(1-s)+ \eta_2 s$ and re-express \eqref{V} in complex coordinates $r^2= z \Bar{z}$:
\begin{equation}
     J= |\tilde{Q}|\int_0^1 ds \frac{z \bar{z}}{1- z\bar{z}}
\end{equation}
Which matches the spin chain computation precisely. We can evaluate the integral explicitly by noting that there is a simple relation between the angular momentum density inside and outside the droplet,
\begin{equation}
    \frac{z \bar{z}}{1- z\bar{z}}= \frac{1}{1-z\bar{z}}-1
\end{equation}
Which reduces to the same integral \eqref{Jintegral} as before:

\begin{equation}
    J+ |Q|=  -|Q| \frac{\arctan\left(\frac{|\eta_2|^2-|\eta_1|^2+|\eta_1-\eta_2|^2}{2\sqrt{|\eta_1 \times \eta_2|^2-|\eta_1-\eta_2|^2}}\right)-\arctan\left(\frac{|\eta_2|^2-|\eta_1|^2-|\eta_1-\eta_2|^2}{2\sqrt{|\eta_1 \times \eta_2|^2-|\eta_1-\eta_2|^2}}\right)}{\sqrt{|\eta_1 \times \eta_2|^2-|\eta_1-\eta_2|^2}}+ \dots
\end{equation}
\subsection{Magnon S-matrix and Bethe Ansatz}
An interesting property of the coherent state ansatz for the one-loop Hamiltonian is that it leads to solutions to the Bethe equations. To see this more explicitly, substituting the coherent state ansatz into the Hamiltonian and minimizing over the complex parameters $z_i$ one finds that their second difference vanishes:

\begin{equation}\label{2nddiff}
    z_{i+1}- 2 z_i+ z_{i-1}=0
\end{equation}
We can always choose to parametrize the complex variables $z_i= e^{\rho_i}$, which leads to the relation:

\begin{equation}
    e^{\rho_{i+1}-\rho_{i-1}}-2 e^{\rho_i-\rho_{i-1}}+1=0
\end{equation}
To make the connection to the Bethe ansatz more explicit it is convenient to make a change of variables:

\begin{equation}
\begin{aligned}
       ip_{l+1}+ip_l&= \rho_{l+1}-\rho_{l-1}\\
       ip_{l}&= \rho_l-\rho_{l-1}\\
\end{aligned}
\end{equation}
Solving these relations leads to an expression purely in terms of $p_{1,2}$,

\begin{equation}
    e^{ip_2+ip_1}-2 e^{ip_1}+1=0
\end{equation}
which can be recognized as a pole for the 2-magnon S-matrix for the $SU(2)$ sector:
\begin{equation}\label{Smatrix}
\begin{aligned}
     S_{12}^{\mathfrak{su}(2)}&= -\frac{  e^{ip_2+ip_1}-2 e^{ip_2}+1}{  e^{ip_2+ip_1}-2 e^{ip_1}+1}= \frac{u_1-u_2-i}{u_1-u_2+i}\\
       e^{ip_l}&=\frac{u_l-i/2}{u_l+i/2}
\end{aligned}
\end{equation}
The interpretation of this pole is that we have formed a bound state of the magnons. Such magnon bound states have the same dispersion relation of \eqref{charges}, as they are also in short representations of the centrally extended spin chain \cite{Dorey:2006dq}. In this sense, the Bethe ansatz computation and the sigma model are fully consistent with each other.

\subsubsection{Analytic Continuation to $SL(2)$}
Another important fact is that the 2-magnon S-matrix for the $SL(2)$ sector is related up to a phase factor to the inverse of the $SU(2)$ S-matrix
(we follow \cite{Gromov:2005gp}):

\begin{equation}
\begin{aligned}
       S_{12}^{\mathfrak{sl}(2)}&\propto  -\frac{  e^{ip_2+ip_1}-2 e^{ip_1}+1}{  e^{ip_2+ip_1}-2 e^{ip_2}+1}= \frac{u_1-u_2+i}{u_1-u_2-i}\\
       e^{ip_l}&=\frac{u_l+i/2}{u_l-i/2}
\end{aligned}
\end{equation}
In the formula above, the right hand side of the S-matrix looks the same, but the identification of momentum with the $u$ variable differs, and is clearly the inverse of the one of $SU(2)$ above.

In particular, the role of poles and zeros is exchanged with respect to the $SU(2)$ sector. Naively one would expect that the Cuntz oscillator representation of the $SU(2)$ Hamiltonian can be analytically continued by allowing the complex parameters $z_i$ to lie outside the unit disk, but this is not the case because then the ground state is no longer normalizable and the S-matrix would not have the correct pole structure. In particular the relation \eqref{2nddiff} would lead to a zero of the 2 magnon S-matrix rather than a pole. If instead one exchanges $z_i \leftrightarrow\frac{1}{z_i}=\tilde{z_i}$, one finds that the zeros of \eqref{Smatrix} are exchanged with poles, while having $|\tilde{z_i}|>1$. Substituting this directly in \eqref{Rcharge} leads to the expression:

\begin{equation}
    J = \sum_{i=1}^k \frac{1}{z_i\Bar{z}_i-1}
\end{equation}
where $k$ now counts the number of derivatives between the $Z$ operators as opposed to counting the number of $Y$ insertions between $Z$'s. When we take the continuum limit of this expression one obtains the same integral that appears in the the string sigma model computation for $AdS_3\times S^1$. \\
This analytic continuation also appears naturally in the LLM coordinates as the transformation that maps the inside and outside of a droplet to each other. It now also appears as a consistency condition for the analytically continued S-matrix. The fact that the string solutions are straight lines on the LLM plane would  translate to having a pole on the magnon S-matrix for the $SL(2)$ sector.

\section{Discussion}
In this work we studied a class of  open string solutions ending on (dual) giant gravitons in $\frac{1}{2}$-BPS geometries of type IIB string theory.  We showed that important simplifications happen when one takes into consideration the appropriate boundary conditions for the end points of the string. The solutions found have a relativistic dispersion relation so that they generalize the giant magnon solutions to open strings in more general backgrounds. This dispersion relation is related to a shortening condition of the central
charge extension symmetry on the worldsheet.
In the solutions we studied, the strings are allowed to extend into the non-compact dimensions of the spacetime, and they have well-defined finite charges inside  droplet regions in the LLM plane. We also found that the solutions cannot be extended between regions of different colors in the LLM plane without having divergences in the approximate charges that generate translations parallel to the droplets, or in the case of concentric geometries the angular momentum $J$ associated to rotations in the LLM plane. Additionally, the coordinates of the string along the non-vanishing three-sphere fiber directions are related to the pull-back of a one form $V$ of the geometry, so that the string charge density  on the fiber (and the analytic solution itself) diverges at the boundaries of the droplets. As a consequence, one can expect that the operators corresponding to such crossing string solutions do not exist within the class considered.
 These divergences are also suggestive of possible instabilities for the states corresponding to the solutions where the strings grace a droplet, like when one crosses a wall of marginal stability ( see also \cite{Berenstein:2014zxa}). Less supersymmetric solutions are allowed if the strings leave the LLM plane.\\

Finally, although one can explictly match the sigma model answer for $J$ to a simple computation in the $SU(2)$ sector of the dual spin chain description, the analogous computation for the $SL(2)$ has not been carried out and seems quite difficult. The fact that the sigma model predicts that the central charge is a constant per unit worldsheet angular momentum (charge) appears in the dual $SU(2)$ spin chain description as a condition on the coherent state ansatz for the excited states. These conditions on the parameters $z_l= e^{\rho_l }$ can be related to a Bethe ansatz solution to the Heisenberg spin chain, where the ratios $z_l/z_{l+1}$ behave like Bethe roots.

 It is natural for us to expect that a similar condition should arise in the $SL(2)$ sector, as the sigma model descriptions of both sectors are analytic continuations of each other. This Bethe ansatz analysis was carried out by us in this paper and gives rise to a consistency check with the sigma model computation.
The analysis is interesting in that the variables that would naturally play the role of $z_i$, which are required to satisfy $|z_i|<1$ on the sphere need to be transformed to $z_i\to 1/z_i$ so that now they are lying outside the LLM plane. This is exactly what is required to exchange the poles and zeros in the S-matrix, which is a natural relation between the $SL(2)$ and $SL(2)$ Bethe ansatz S-matrices.

Similar bubbling solutions exist for six dimensional supergravity \cite{Martelli:2004xq,Liu:2004hy} but the details of the geometries are more complicated. Our results seem to apply to an isolated class of these geometries, at least when the torus fibration of the geometries is a product of circles. The analysis for general 6d $\frac{1}{2}$ BPS bubbling geometries is complicated by the appearance of an axion corresponding to the off-diagonal component of the metric for the torus fiber, which introduces additional singularities where non-trivial sectional circles of the torus vanish. Similarly, when we try to work with $\frac{1}{4}$ and $\frac{1}{8}$ BPS geometries in ten dimensions,  the flat LLM plane is replaced by a four or six dimensional K{\"a}hler base with three or five dimensional droplets \cite{Chen:2007du}, and constructing explicit metrics is a very non-trivial task.
In all these situations the backgrounds preserve only $8$ or $4$ supersymmetries rather than 16 and this seems to be the main reason the analysis is more complicated.
Even open strings suspended between AdS giants in $AdS_5 \times S^5$ that together only preserve $\frac{1}{4}$ of the supersymmetries seem to have additional corrections.  We are currently investigating these issues. These might be more tractable in the limit that the AdS giant gravitons carry very large charge, as in this limit the analysis also simplified for the $\frac{1}{2}$ BPS case.

\acknowledgments

The work of D.B. is supported in part by the Department of Energy under grant DE-SC 0011702.


\begin{thebibliography}{99}

\bibitem{Maldacena:1997re}
J.~M.~Maldacena,
``The Large N limit of superconformal field theories and supergravity,''
Int. J. Theor. Phys. \textbf{38}, 1113-1133 (1999)
doi:10.1023/A:1026654312961
[arXiv:hep-th/9711200 [hep-th]].

\bibitem{Beisert:2010jr}
N.~Beisert, C.~Ahn, L.~F.~Alday, Z.~Bajnok, J.~M.~Drummond, L.~Freyhult, N.~Gromov, R.~A.~Janik, V.~Kazakov, T.~Klose, G.~P.~Korchemsky, C.~Kristjansen, M.~Magro, T.~McLoughlin, J.~A.~Minahan, R.~I.~Nepomechie, A.~Rej, R.~Roiban, S.~Schafer-Nameki, C.~Sieg, M.~Staudacher, A.~Torrielli, A.~A.~Tseytlin, P.~Vieira, D.~Volin and K.~Zoubos,
``Review of AdS/CFT Integrability: An Overview,''
Lett. Math. Phys. \textbf{99}, 3-32 (2012)
doi:10.1007/s11005-011-0529-2
[arXiv:1012.3982 [hep-th]].

\bibitem{Minahan:2002ve}
J.~A.~Minahan and K.~Zarembo,
``The Bethe ansatz for N=4 superYang-Mills,''
JHEP \textbf{03}, 013 (2003)
doi:10.1088/1126-6708/2003/03/013
[arXiv:hep-th/0212208 [hep-th]].

\bibitem{Beisert:2003yb}
N.~Beisert and M.~Staudacher,
``The N=4 SYM integrable super spin chain,''
Nucl. Phys. B \textbf{670}, 439-463 (2003)
doi:10.1016/j.nuclphysb.2003.08.015
[arXiv:hep-th/0307042 [hep-th]].

\bibitem{Beisert:2005tm}
N.~Beisert,
``The SU(2|2) dynamic S-matrix,''
Adv. Theor. Math. Phys. \textbf{12}, 945-979 (2008)
doi:10.4310/ATMP.2008.v12.n5.a1
[arXiv:hep-th/0511082 [hep-th]].

\bibitem{Hofman:2006xt}
D.~M.~Hofman and J.~M.~Maldacena,
``Giant Magnons,''
J. Phys. A \textbf{39}, 13095-13118 (2006)
doi:10.1088/0305-4470/39/41/S17
[arXiv:hep-th/0604135 [hep-th]].

\bibitem{Berenstein:2002jq}
D.~E.~Berenstein, J.~M.~Maldacena and H.~S.~Nastase,
``Strings in flat space and pp waves from N=4 superYang-Mills,''
JHEP \textbf{04}, 013 (2002)
doi:10.1088/1126-6708/2002/04/013
[arXiv:hep-th/0202021 [hep-th]].

\bibitem{Berenstein:2013md}
D.~Berenstein,
``Giant gravitons: a collective coordinate approach,''
Phys. Rev. D \textbf{87}, no.12, 126009 (2013)
doi:10.1103/PhysRevD.87.126009
[arXiv:1301.3519 [hep-th]].


\bibitem{Berenstein:2013eya}
D.~Berenstein and E.~Dzienkowski,
``Open spin chains for giant gravitons and relativity,''
JHEP \textbf{08}, 047 (2013)
doi:10.1007/JHEP08(2013)047
[arXiv:1305.2394 [hep-th]].

\bibitem{Berenstein:2014isa}
D.~Berenstein and E.~Dzienkowski,
``Giant gravitons and the emergence of geometric limits in beta-deformations of $ \mathcal{N}=4 $ SYM,''
JHEP \textbf{01}, 126 (2015)
doi:10.1007/JHEP01(2015)126
[arXiv:1408.3620 [hep-th]].

\bibitem{Berenstein:2014zxa}
D.~Berenstein,
``On the central charge extension of the $ \mathcal{N}=4 $ SYM spin chain,''
JHEP \textbf{05}, 129 (2015)
doi:10.1007/JHEP05(2015)129
[arXiv:1411.5921 [hep-th]].

\bibitem{Dzienkowski:2015zba}
E.~Dzienkowski,
``Excited States of Open Strings From $\mathcal{N}=4$ SYM,''
JHEP \textbf{12}, 036 (2015)
doi:10.1007/JHEP12(2015)036
[arXiv:1507.01595 [hep-th]].


\bibitem{deCarvalho:2020pdp}
S.~de Carvalho, R.~de Mello Koch and M.~Kim,
``Central Charges for the Double Coset,''
JHEP \textbf{05}, 007 (2020)
doi:10.1007/JHEP05(2020)007
[arXiv:2001.10181 [hep-th]].

\bibitem{Berenstein:2020grg}
D.~Berenstein and A.~Holguin,
``Open giant magnons suspended between dual giant gravitons in ${\cal N}=4$ SYM,''
[arXiv:2006.08649 [hep-th]].


\bibitem{McGreevy:2000cw}
J.~McGreevy, L.~Susskind and N.~Toumbas,
``Invasion of the giant gravitons from Anti-de Sitter space,''
JHEP \textbf{06}, 008 (2000)
doi:10.1088/1126-6708/2000/06/008
[arXiv:hep-th/0003075 [hep-th]].

\bibitem{Grisaru:2000zn}
M.~T.~Grisaru, R.~C.~Myers and O.~Tafjord,
``SUSY and goliath,''
JHEP \textbf{08}, 040 (2000)
doi:10.1088/1126-6708/2000/08/040
[arXiv:hep-th/0008015 [hep-th]].

\bibitem{Hashimoto:2000zp}
A.~Hashimoto, S.~Hirano and N.~Itzhaki,
``Large branes in AdS and their field theory dual,''
JHEP \textbf{08}, 051 (2000)
doi:10.1088/1126-6708/2000/08/051
[arXiv:hep-th/0008016 [hep-th]].

\bibitem{Berenstein:2004kk}
D.~Berenstein,
``A Toy model for the AdS / CFT correspondence,''
JHEP \textbf{07}, 018 (2004)
doi:10.1088/1126-6708/2004/07/018
[arXiv:hep-th/0403110 [hep-th]].

\bibitem{Corley:2001zk}
S.~Corley, A.~Jevicki and S.~Ramgoolam,
``Exact correlators of giant gravitons from dual N=4 SYM theory,''
Adv. Theor. Math. Phys. \textbf{5}, 809-839 (2002)
doi:10.4310/ATMP.2001.v5.n4.a6
[arXiv:hep-th/0111222 [hep-th]].

\bibitem{Balasubramanian:2001nh}
V.~Balasubramanian, M.~Berkooz, A.~Naqvi and M.~J.~Strassler,
``Giant gravitons in conformal field theory,''
JHEP \textbf{04}, 034 (2002)
doi:10.1088/1126-6708/2002/04/034
[arXiv:hep-th/0107119 [hep-th]].


\bibitem{Susskind:1998dq}
L.~Susskind and E.~Witten,
``The Holographic bound in anti-de Sitter space,''
[arXiv:hep-th/9805114 [hep-th]].

\bibitem{deBoer:1999tgo}
J.~de Boer, E.~P.~Verlinde and H.~L.~Verlinde,
``On the holographic renormalization group,''
JHEP \textbf{08}, 003 (2000)
doi:10.1088/1126-6708/2000/08/003
[arXiv:hep-th/9912012 [hep-th]].


\bibitem{Gubser:2002tv}
S.~S.~Gubser, I.~R.~Klebanov and A.~M.~Polyakov,
``A Semiclassical limit of the gauge / string correspondence,''
Nucl. Phys. B \textbf{636}, 99-114 (2002)
doi:10.1016/S0550-3213(02)00373-5
[arXiv:hep-th/0204051 [hep-th]].

\bibitem{Frolov:2003qc}
S.~Frolov and A.~A.~Tseytlin,
``Multispin string solutions in AdS(5) x S**5,''
Nucl. Phys. B \textbf{668}, 77-110 (2003)
doi:10.1016/S0550-3213(03)00580-7
[arXiv:hep-th/0304255 [hep-th]].

\bibitem{Kruczenski:2004wg}
M.~Kruczenski,
``Spiky strings and single trace operators in gauge theories,''
JHEP \textbf{08}, 014 (2005)
doi:10.1088/1126-6708/2005/08/014
[arXiv:hep-th/0410226 [hep-th]].


\bibitem{Alday:2007mf}
L.~F.~Alday and J.~M.~Maldacena,
``Comments on operators with large spin,''
JHEP \textbf{11}, 019 (2007)
doi:10.1088/1126-6708/2007/11/019
[arXiv:0708.0672 [hep-th]].

\bibitem{Chen:2007gh}
H.~Y.~Chen, D.~H.~Correa and G.~A.~Silva,
``Geometry and topology of bubble solutions from gauge theory,''
Phys. Rev. D \textbf{76}, 026003 (2007)
doi:10.1103/PhysRevD.76.026003
[arXiv:hep-th/0703068 [hep-th]].




\bibitem{Koch:2016jnm}
R.~de Mello Koch, C.~Mathwin and H.~J.~R.~van Zyl,
``LLM Magnons,''
JHEP \textbf{03}, 110 (2016)
doi:10.1007/JHEP03(2016)110
[arXiv:1601.06914 [hep-th]].



\bibitem{Lin:2004nb}
H.~Lin, O.~Lunin and J.~M.~Maldacena,
``Bubbling AdS space and 1/2 BPS geometries,''
JHEP \textbf{10}, 025 (2004)
doi:10.1088/1126-6708/2004/10/025
[arXiv:hep-th/0409174 [hep-th]].

\bibitem{Koch:2016qkp}
R.~de Mello Koch and H.~J.~R.~van Zyl,
``Inelastic Magnon Scattering,''
Phys. Lett. B \textbf{768}, 187-191 (2017)
doi:10.1016/j.physletb.2017.02.056
[arXiv:1603.06414 [hep-th]].


\bibitem{Berenstein:2005fa}
D.~Berenstein, D.~H.~Correa and S.~E.~Vazquez,
``Quantizing open spin chains with variable length: An Example from giant gravitons,''
Phys. Rev. Lett. \textbf{95}, 191601 (2005)
doi:10.1103/PhysRevLett.95.191601
[arXiv:hep-th/0502172 [hep-th]].





\bibitem{Correa:2006yu}
D.~H.~Correa and G.~A.~Silva,
``Dilatation operator and the super Yang-Mills duals of open strings on AdS giant gravitons,''
JHEP \textbf{11}, 059 (2006)
doi:10.1088/1126-6708/2006/11/059
[arXiv:hep-th/0608128 [hep-th]].




\bibitem{Minahan:2006bd}
J.~A.~Minahan, A.~Tirziu and A.~A.~Tseytlin,
``Infinite spin limit of semiclassical string states,''
JHEP \textbf{08}, 049 (2006)
doi:10.1088/1126-6708/2006/08/049
[arXiv:hep-th/0606145 [hep-th]].




\bibitem{Berenstein:2007zf}
D.~Berenstein and S.~E.~Vazquez,
``Giant magnon bound states from strongly coupled N=4 SYM,''
Phys. Rev. D \textbf{77}, 026005 (2008)
doi:10.1103/PhysRevD.77.026005
[arXiv:0707.4669 [hep-th]].

\bibitem{Balasubramanian:2004nb}
V.~Balasubramanian, D.~Berenstein, B.~Feng and M.~x.~Huang,
``D-branes in Yang-Mills theory and emergent gauge symmetry,''
JHEP \textbf{03}, 006 (2005)
doi:10.1088/1126-6708/2005/03/006
[arXiv:hep-th/0411205 [hep-th]].


\bibitem{deMelloKoch:2007rqf}
R.~de Mello Koch, J.~Smolic and M.~Smolic,
``Giant Gravitons - with Strings Attached (I),''
JHEP \textbf{06}, 074 (2007)
doi:10.1088/1126-6708/2007/06/074
[arXiv:hep-th/0701066 [hep-th]].

\bibitem{deMelloKoch:2007nbd}
R.~de Mello Koch, J.~Smolic and M.~Smolic,
``Giant Gravitons - with Strings Attached (II),''
JHEP \textbf{09}, 049 (2007)
doi:10.1088/1126-6708/2007/09/049
[arXiv:hep-th/0701067 [hep-th]].

\bibitem{Bekker:2007ea}
D.~Bekker, R.~de Mello Koch and M.~Stephanou,
``Giant Gravitons - with Strings Attached. III.,''
JHEP \textbf{02}, 029 (2008)
doi:10.1088/1126-6708/2008/02/029
[arXiv:0710.5372 [hep-th]].

\bibitem{Beisert:2003jj}
N.~Beisert,
``The complete one loop dilatation operator of N=4 superYang-Mills theory,''
Nucl. Phys. B \textbf{676}, 3-42 (2004)
doi:10.1016/j.nuclphysb.2003.10.019
[arXiv:hep-th/0307015 [hep-th]].

\bibitem{Dorey:2006dq}
N.~Dorey,
``Magnon Bound States and the AdS/CFT Correspondence,''
J. Phys. A \textbf{39}, 13119-13128 (2006)
doi:10.1088/0305-4470/39/41/S18
[arXiv:hep-th/0604175 [hep-th]].


\bibitem{Gromov:2005gp}
N.~Gromov and V.~Kazakov,
``Double scaling and finite size corrections in sl(2) spin chain,''
Nucl. Phys. B \textbf{736}, 199-224 (2006)
doi:10.1016/j.nuclphysb.2005.12.006
[arXiv:hep-th/0510194 [hep-th]].






\bibitem{Martelli:2004xq}
D.~Martelli and J.~F.~Morales,
``Bubbling AdS(3),''
JHEP \textbf{02}, 048 (2005)
doi:10.1088/1126-6708/2005/02/048
[arXiv:hep-th/0412136 [hep-th]].

\bibitem{Liu:2004hy}
J.~T.~Liu and D.~Vaman,
``Bubbling 1/2 BPS solutions of minimal six-dimensional supergravity,''
Phys. Lett. B \textbf{642}, 411-419 (2006)
doi:10.1016/j.physletb.2006.09.059
[arXiv:hep-th/0412242 [hep-th]].

\bibitem{Chen:2007du}
B.~Chen, S.~Cremonini, A.~Donos, F.~L.~Lin, H.~Lin, J.~T.~Liu, D.~Vaman and W.~Y.~Wen,
``Bubbling AdS and droplet descriptions of BPS geometries in IIB supergravity,''
JHEP \textbf{10}, 003 (2007)
doi:10.1088/1126-6708/2007/10/003
[arXiv:0704.2233 [hep-th]].














\bibitem{Lin:2010sba}
H.~Lin, A.~Morisse and J.~P.~Shock,
JHEP \textbf{06}, 055 (2010)
doi:10.1007/JHEP06(2010)055
[arXiv:1003.4190 [hep-th]].




\end{thebibliography}
\end{document}